\documentclass[pdflatex,sn-mathphys-num]{sn-jnl}

\usepackage{graphicx}%
\usepackage{multirow}%
\usepackage{amsmath,amssymb,amsfonts}%
\usepackage{amsthm}%
\usepackage{mathrsfs}%
\usepackage[title]{appendix}%
\usepackage{xcolor}%
\usepackage{textcomp}%
\usepackage{manyfoot}%
\usepackage{booktabs}%
\usepackage{algorithm}%
\usepackage{algorithmicx}%
\usepackage{algpseudocode}%
\usepackage{listings}%

\theoremstyle{thmstyleone}%
\theoremstyle{thmstyletwo}%

\theoremstyle{thmstylethree}%

\begin{document}

\title[Article Title]{SuperSalt: Equivariant Neural Network Force Fields for Multicomponent Molten Salts System}


\author*[1]{\fnm{Chen} \sur{Shen}}\email{cshen89@wisc.edu}
\equalcont{These authors contributed equally to this work.}

\author[1]{\fnm{Siamak} \sur{Attarian}}\email{sattarian@wisc.edu}
\equalcont{These authors contributed equally to this work.}

\author[2]{\fnm{Yixuan} \sur{Zhang}}
\email{yixuan.zhang@tu-darmstadt.de}


\author[2]{\fnm{Hongbin} \sur{Zhang}}
\email{hzhang@tmm.tu-darmstadt.de}

\author[3]{\fnm{Mark} \sur{Asta}}
\email{mdasta@berkeley.edu}

\author[1]{\fnm{Izabela} \sur{Szlufarska}}\email{szlufarska@wisc.edu}

\author*[1]{\fnm{Dane} \sur{Morgan}}\email{ddmorgan@wisc.edu}

\affil[1]{\orgdiv{Materials Science and Engineering}, \orgname{University of Wisconsin-Madison}, \orgaddress{\city{Madison}, \postcode{53706}, \state{Wisconsin}, \country{United States}}}

\affil[2]{\orgdiv{Materials Science}, \orgname{Technical University of Darmstadt}, \orgaddress{\city{Darmstadt}, \postcode{64287}, \state{Hessen}, \country{Germany}}}

\affil[3]{\orgdiv{Materials Science and Engineering}, \orgname{University of California}, \orgaddress{ \city{Berkeley}, \postcode{94720}, \state{California}, \country{United States}}}


\abstract{Molten salts are crucial for clean energy applications, yet exploring their thermophysical properties across diverse chemical space remains challenging. We present the development of a machine learning interatomic potential (MLIP) called SuperSalt, which targets 11-cation chloride melts and captures the essential physics of molten salts with near-DFT accuracy. Using an efficient workflow that integrates systems of one, two, and 11 components, the SuperSalt potential can accurately predict thermophysical properties such as density, bulk modulus, thermal expansion, and heat capacity. Our model is validated across a broad chemical space, demonstrating excellent transferability. We further illustrate how Bayesian optimization combined with SuperSalt can accelerate the discovery of optimal salt compositions with desired properties. This work provides a foundation for future studies that allows easy extensions to more complex systems, such as those containing additional elements. SuperSalt represents a shift towards a more universal, efficient, and accurate modeling of molten salts for advanced energy applications.}


\maketitle

\section{Introduction}\label{sec1}
Over the past several decades, atomic-scale simulation has emerged as an indispensable tool for predicting and providing microscopic insights into experimentally observed phenomena in molten salt materials. Numerous scientifically important thermophysical and transport properties of molten salts can be accurately evaluated using molecular dynamics (MD) simulations, in which atomic motion is determined by integrating Newton’s second law of motion. The predictive accuracy of MD simulations is highly dependent on the quality of the underlying potential energy surface (the potential) that determines the forces that act on each atom. 
Standard physics-based approaches to potentials, such as classical force fields (FFs)~\cite{fumi1964ionic,sangster1976interionic,aguado2003multipoles,salanne2012including} and quantum mechanics (QM) methods, particularly density functional theory (DFT)~\cite{hohenberg1964inhomogeneous,kohn1965self}, have been employed successfully to model various molten salt systems~\cite{bengtson2014first,nam2015redox,li2017dynamic}. 
However, these approaches involve a well-known trade-off between computational cost, accuracy, and generality. Classical FFs, while computationally efficient, often require reparameterization for specific systems or reactions. Conversely, QM-based methods are computationally expensive, limiting their utility for large-scale and long-timescale MD simulations. 
This trade-off is especially critical for modeling complicated properties, such as viscosities. Thus, MD simulations urgently need a fast, accurate, and broadly applicable reactive potential to enable the reliable prediction of thermophysical salt properties and salt reactions with their environment (e.g., corrosion). 

Recently, machine learning interatomic potentials (MLIPs)~\cite{behler2007generalized} have emerged as a promising alternative to overcome these limitations. MLIPs can achieve near-DFT accuracy while maintaining computational costs close to those of classical FFs, enabling MD simulations with $10^3$–$10^4$ atoms over nanosecond timescales. This capability unlocks unprecedented opportunities for studying complex molten salt systems~\cite{lam2021modeling}. 
Up to this point almost all MLIPs for molten salts were tailored to specific chemical compositions, focusing on a narrow range of properties. For example, Attarian et al. developed MLIPs for FLiBe molten salts~\cite{attarian2024studies,attarian2022thermophysical}, accurately predicting properties such as density, heat capacity, and ionic conductivity. Similarly, Lu et al. constructed MLIPs for SrCl$_2$–KCl–MgCl$_2$ melts~\cite{zhao2024deep}, predicting local structures and thermophysical properties. This has led to 38 separate papers since 2021 fitting MLIPs to chloride salts (mainly on unary and binary systems). While effective, these approaches necessitate repeated efforts to construct training datasets and fit MLIPs for each new system, a time-intensive process. 

A promising alternative lies in the development of
so-called universal potentials~\cite{batatia2023foundation,chen2022universal,choudhary2023unified,takamoto2022towards,xie2024gptff,merchant2023scaling,park2024scalable,deng2023chgnet}, which are trained on datasets encompassing 50 or more chemical species. These potentials offer broad applicability and reduce the need for system-specific reparameterization. However, they often exhibit significant errors in some properties and are generally less accurate than MLIPs designed for specific systems. For instance, M3GNet demonstrates relatively high errors in predicting vibrational properties~\cite{yu2024systematic}.

To address these challenges, we propose an intermediate approach that focuses on a subset of 10–20 chemically similar elements, aiming to achieve near-DFT accuracy while maintaining computational efficiency. Such a strategy reduces the inefficiencies of developing numerous highly specific MLIPs while enabling high accuracy across a substantial composition space. Notably, a single MLIP trained on $N$ elements effectively models all $2^N$ suballoys, offering substantial computational savings. For example, a potential trained on 11 elements inherently describes all 2048 suballoys, significantly simplifying the exploration of complex chemical spaces.

In this work, we developed an efficient workflow to construct an MLIP, referred to as the "SuperSalt" potential, tailored to all liquid-phase compositions of 11-cation chloride melts: LiCl, NaCl, KCl, RbCl, CsCl, MgCl$_2$, CaCl$_2$, SrCl$_2$, BaCl$_2$, ZnCl$_2$, and ZrCl$_4$. Extensive validation demonstrates that the SuperSalt potential achieves near-DFT accuracy in predicting key properties such as densities, bulk moduli, radial distribution functions, specific heat capacities, and thermal expansion coefficients for multicomponent molten salt systems. Furthermore, the potential exhibits excellent transferability across different chemical compositions, enabling reliable predictions for unexplored systems.
The SuperSalt potential also facilitates targeted optimization of material properties within the 11-cation composition space. By leveraging Bayesian optimization (BO)~\cite{frazier2016bayesian,lookman2019active}, we identified optimal compositions that meet specific property requirements, showcasing the potential's utility in accelerating materials discovery. Such discovery is not possible with traditional empirical or ab initio methods and is enabled by our ability to model 11 cations with machine learning-based strategies. The SuperSalt potential offers a robust and efficient platform for simulating and optimizing multicomponent molten salt systems. With further extension, this approach has the potential to provide a foundational MLIP framework for the majority of molten salts of interest, bridging the gap between computational efficiency, accuracy, and general applicability.

This paper is arranged as follows: The results section first outlines the comprehensive workflow developed for generating a robust SuperSalt potential, including details on the training dataset and the active learning framework employed. Secondly, we present the training and testing results, showcasing the accuracy and transferability of the SuperSalt model across various molten salt systems. Then, we evaluate the performance of the SuperSalt model by comparing its predictions of thermophysical properties with AIMD and experimental data. The last subsection demonstrates how the SuperSalt potential, in combination with Bayesian optimization, can expedite the discovery of molten salt compositions with optimal properties. Finally, the discussion section highlights the implications of this work for the broader molten salts community and potential future directions for expanding the model.

\section{Results}\label{sec2}
\subsection*{Comprehensive workflow for complex molten salts}
\begin{figure}[h!]
    \centering
    \includegraphics[width=1\textwidth]{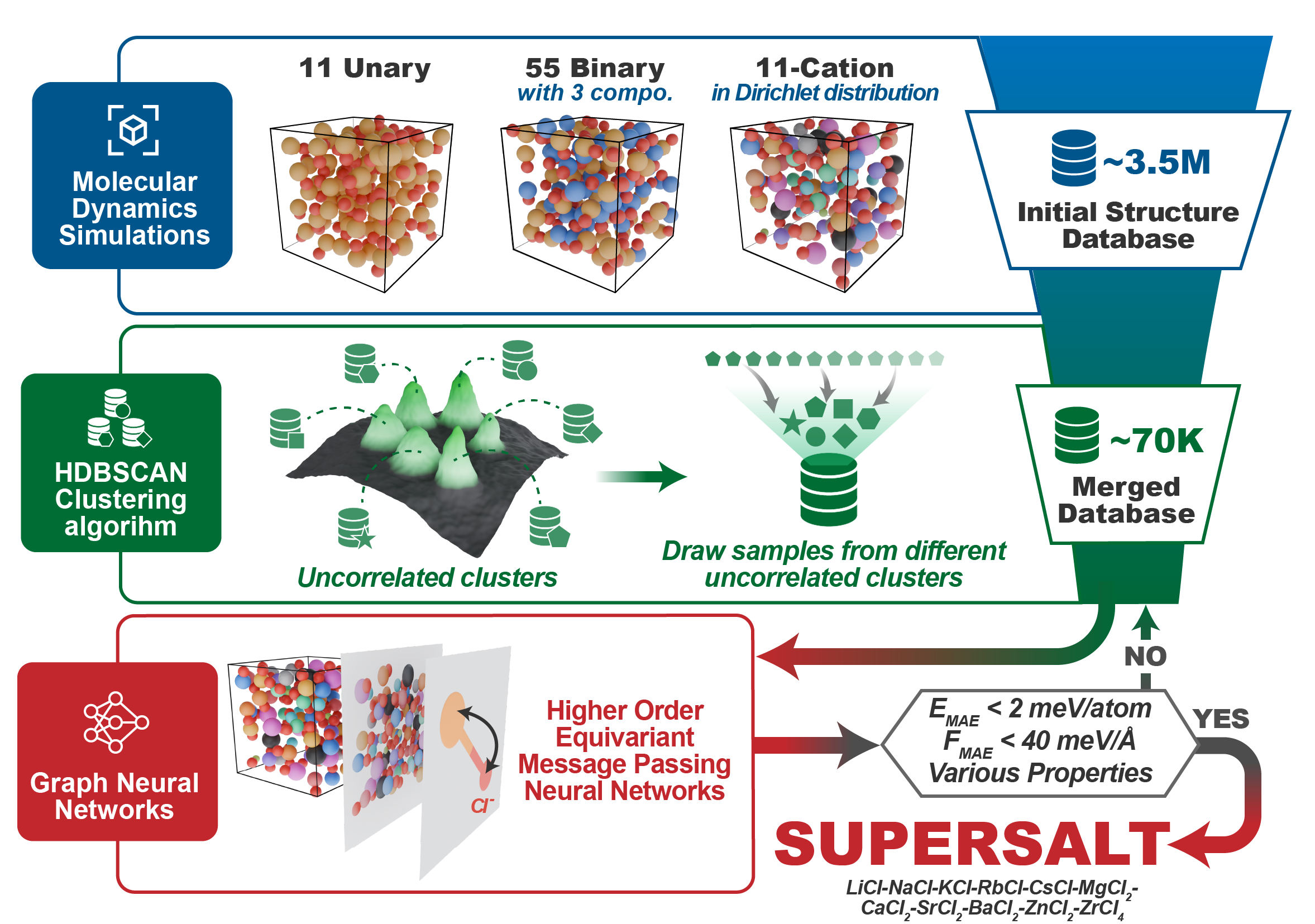}
    \caption{The procedure for obtaining the database and training the model is outlined as follows: 1-component, 2-component, and 11-component compositions are introduced into MD simulations under specific conditions, resulting in the generation of an initial structure database. Subsequently, an active learning algorithm based on HDBSCAN is utilized to automatically sample uncorrelated learning configurations from the initial datasets by leveraging the underlying cluster structure present in the database. The selected learning configurations are sequentially sampled from these uncorrelated clusters and iteratively trained using the MACE model until the desired accuracy is attained.}
    \label{workflow}
\end{figure}

The design of a robust MLIP depends crucially on the choice of training dataset.
For molten salt applications, the primary focus is on liquid properties, such as the heat capacity ($C_P$), thermal expansion, density ($\rho$), molecular structure, equation of state, etc. 
For transferability, the MLIP should be able to make good predictions on random compositions across the whole chemical space of the 11-cation chloride melts. To this end, we have developed a comprehensive dataset of atomistic structures and quantum-mechanical reference data for 11-cation chloride melts, as well as an interatomic potential fitted to that database in the higher-order equivariant message passing neural networks framework (as shown in Figure~\ref{workflow}). 
The chemical space for 11 chlorides consists of $2^{11} = 2048$ chemical combinations, including 11 unaries, 55 binaries, 165 ternaries, etc. Constructing a training dataset by enumerating all possible chemical combinations is formidable.
However, it is well-known that the physics of salts is largely dominated by relatively simple pairwise electrostatics, so we expected that the fitting of 1- and 2-component systems would provide much of the essential physics. We assumed some many-component interactions would be important and that the MLIP would require some experience with the many-element environments to ensure transferability, so we also included data from 11-cation chloride salts in our training set.
We employed the Dirichlet distribution method~\cite{ng2011dirichlet} to control and generate the compositions of the 11-component system systematically. This method provides flexibility in sampling different composition spaces while maintaining a fixed total concentration, which is crucial for accurately capturing the complex behavior of the multicomponent molten salt system. By using this approach, we ensure that each configuration sampled represents a physically realistic and consistent composition, facilitating the exploration of various mixtures across the 11-component phase space. Therefore, our training dataset included only 1-component, 2-component, and 11-component configurations. We chose the nonlinear neural-network multilayer atomic cluster expansion (MACE) model~\cite{bochkarev2022multilayer} for the SuperSalt potential because of its capability to capture higher-order atomic interactions with high accuracy and efficiency. MACE is built on an equivariant message-passing neural network framework, which allows it to handle complex atomic geometries by encoding many-body interactions at each layer of the network. MACE has demonstrated the ability to create a universal foundation model that is applicable across essentially the entire periodic table~\cite{batatia2023foundation,kovacs2023mace}.

The procedure for generating the initial data pool, as illustrated in Figure~\ref{workflow}, follows an NPT "melt-quench" MD approach. This method involves heating a randomly packed structure to an elevated temperature and subsequently quenching it to a lower temperature, creating the initial configuration database. Further details of this procedure are provided in the "Methods" section. Then, we use AL to automatically sample uncorrelated learning configurations from the initial datasets by exploiting the underlying cluster structure embedded in the dataset~\cite{dasgupta2008hierarchical} and partitioning them into “N” uncorrelated clusters~\cite{dasgupta2011two}. The learning configurations are sequentially sampled from the uncorrelated clusters and trained using the MACE model until the desired accuracy has been achieved.
For the clustering, we use HDBSCAN~\cite{campello2013density,mcinnes2017hdbscan,mcinnes2017accelerated}, a density-based hierarchical clustering method. This algorithm has been successfully applied previously to partition and analyze MD trajectories~\cite{melvin2016uncovering}.
The details of the AL workflow and the MD simulations are described in the “Methods” section. Note that the entire "melt-quench" region was mapped to the SuperSalt model with only 2\% of configuration space (initial structures).

The final database was obtained by merging the AL-extracted data of unary, binary, and 11-cation systems, and each subsystem has a few hundred structures. This database contains around 70,000 structures with a total of $\approx$ 7 million atoms. For validation, we held out 10\% of these structures from training, selected at random.

\subsection*{Training and testing results}
Using the AL-learned training dataset, we trained a MACE model (see Method for details on the hyperparameters) using the higher-order equivariant message passing neural networks framework. We refer to this MACE model as the SuperSalt potential, representing the first attempt at constructing a unified MACE model for the multi-component molten salt system.

\begin{figure}[h!]
    \centering
    \includegraphics[width=\textwidth]{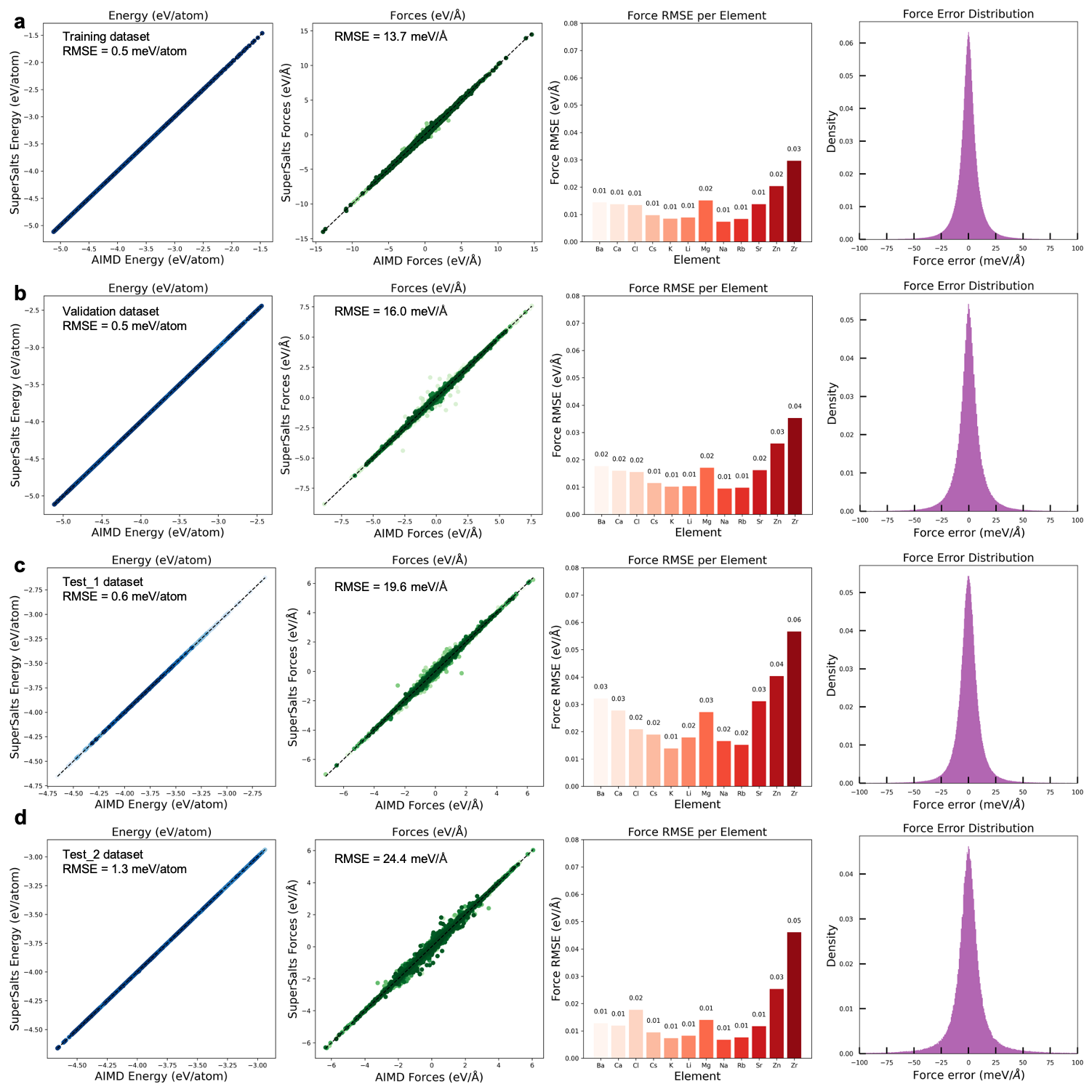}
    \caption{Parity plots for energy and force, comparing DFT reference data with SuperSalt potential predictions, are shown for (a) the training dataset, (b) the validation dataset, (c) Test\_1 containing the 3300-ternary dataset, and (d) Test\_2 containing the 800-multicomponent dataset. The corresponding force RMSE per element for each dataset is presented as histograms in the third column. The testing errors in atomic forces for the four databases in the right panels largely follow a Gaussian distribution and are concentrated in the range of -25 to 25 meV/\AA.
    }
    \label{fig:force}
\end{figure}

The parity plots for energy and force for different databases affirm the high accuracy of the SuperSalt potential (Figure~\ref{fig:force}a-c). 
In Figure~\ref{fig:force}(a and b), for training and validation sets, the deviation of the SuperSalt potential predictions from DFT results for potential energy and atomic forces is reported. 
We use root mean square error (RMSE) as a measure of the overall testing errors. Specifically, the RMSE of total potential energy predictions reaches a value as low as 0.5 meV/atom for both datasets, and the RMSEs of atomic forces predictions are 13.7 meV/\AA and 16.0 meV/\AA, respectively. These very small testing errors suggest an accuracy close to the DFT level.
To further validate the accuracy and check the transferability of the SuperSalt potential, we show its performance on two newly generated testing sets: the Test\_1 set containing 3300-ternary atomic configurations and Test\_2 set containing 800-random multicomponent atomic configurations, which are unrelated to our training set. The details of the testing data generation are discussed in the Methods section.
The testing RMSEs of the SuperSalt for the Test\_1 dataset are 0.6 meV/atom and 19.6 meV/\AA. For the Test\_2 dataset, the testing RMSEs of the SuperSalt are 1.3 meV/atom and 24.4 meV/\AA. These results are somewhat higher than the training and validation data but still very low and even better than those reported as training RMSEs in many previous publications for specific melt systems~\cite{li2021development} (Figure~\ref{fig:force}b and c).
Furthermore, the comparison demonstrates that the SuperSalt potential, trained on 1-component, 2-component, and 11-component structures, performs exceptionally well for 3-component, 4-component systems, and beyond, including in the Test\_1 and Tes\_2 data sets. This highlights the excellent transferability of the SuperSalt potential across the entire chemical space of 11-cation chloride melts.
Interestingly, based on the force errors across different elements, those with higher valence states exhibit more significant discrepancies. Notably, Zr with a 4+ valence state shows the most significant force error in all three scenarios shown in Fig.~\ref{fig:force} (see third column of Figure~\ref{fig:force}a-d). This may be due to the more complex chemical environment occurring around higher valent cations, but more studies are needed to be sure.
The testing errors in atomic forces for three databases largely follow a Gaussian distribution and are concentrated in the range from -25 to 25 meV/\AA (see the histogram plot and normal distribution fitting on the right side of Figure~\ref{fig:force} a-d).

Given the existence of multiple universal MLIPs, it is important to establish that a multi-element but specialized MLIP tailored for molten salt systems can provide better results meaningfully. 
To explore this, the foundation model MACE-MP0~\cite{batatia2023foundation} was tried on the current testing databases by including the DFT-D3 correction to make the predictions consistent with the calculated DFT data of the testing sets. As shown in Figure S1, the resulting errors for the Test\_1 and Test\_2 datasets are 55.8 meV/atom and 49.3 meV/atom, respectively, with corresponding force errors of 191.2 meV/\AA~and 172.3 meV/\AA. These values are approximately an order of magnitude larger than those obtained with our SuperSalt potential, demonstrating that the latter has significantly superior accuracy when applied to molten salt systems.

\subsection*{Performance of SuperSalt Potential}
\subsubsection*{EOS and RDFs}
\begin{figure}[h!]
    \centering
    \includegraphics[width=\textwidth]{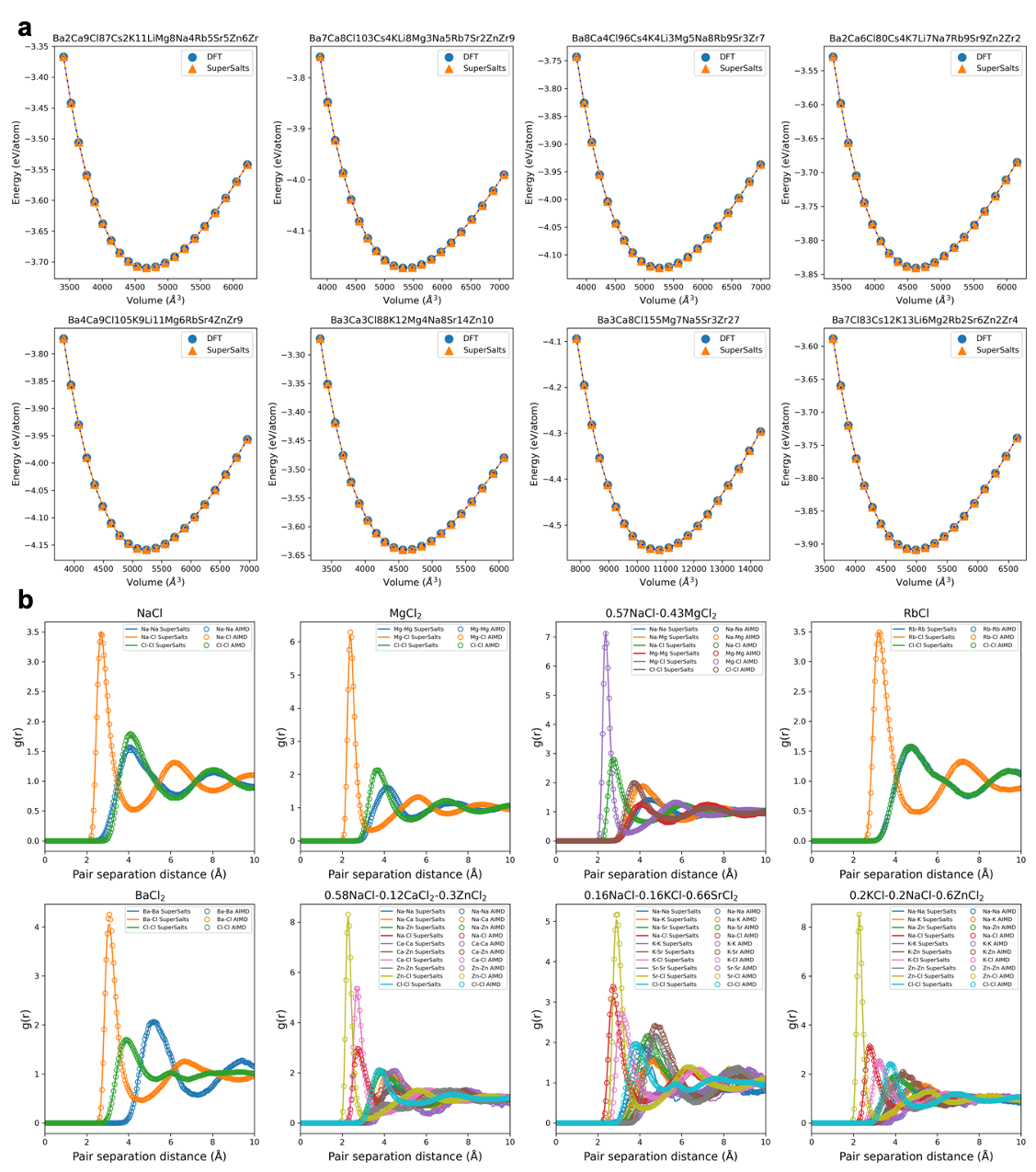}
    \caption{(a) Equations of state (EOS) at 700 K for 8 different multicomponent salts, calculated using SuperSalt-MD, show excellent agreement with AIMD results. (b) Radial distribution functions (RDF) of random molten salts, spanning unary, binary, ternary, and 11-component systems, predicted by SuperSalt-MD, match excellently with those obtained from AIMD.}
    \label{fig:rdf_eos}
\end{figure}

Next, we evaluate the performance of our SuperSalt potential by testing various liquid properties commonly used to identify potential candidate molten salts.
We used random atomic configurations and compared their energy vs. volume curves obtained from SuperSalt potential and DFT. To do this, we started by generating random atomic configurations and ran MD simulations at three temperatures, 700 K, 1200 K, and 1500 K, then randomly picked structures from these simulations and applied hydrostatic strains between -10\% to 10\% at every 1\% point on these structures and calculated their energy by both methods. Overall, 66 structures were selected for this test, and the energy vs. volume curves for some of these structures and partial results are shown in Figure~\ref{fig:rdf_eos}a. From these curves, we calculated the bulk modulus, the minimum energy, and the equilibrium density for each structure at 0 K. We find that the RMSE in the predictions of bulk modulus, minimum energy, and equilibrium density between the SuperSalt potential and DFT are 0.23 GPa, 2.4 meV/atom, and $3 \times 10^{-3}$ g/cm$^3$,  respectively. The list of the structures and the detailed comparison between the calculated values using DFT and SuperSalt potential are provided in the supplementary materials. 
In the supplementary materials, we also make the same comparison between MACE-MP0 and DFT without considering D3 correction so it would be consistent with the MACE-MP0 training data. The MACE-MP0 results show RMSEs of 2.7 GPa, 22 meV/atom, and $2.7 \times 10^{-2}$ g/cm$^3$, respectively, for bulk modulus, minimum energy, and equilibrium density predictions. Thus, we again find that the SuperSalt potential, which is directly fitted to the molten salt configurations, outperforms the MACE-MP0 in calculating key properties.

We then examine the liquid structures from both our AIMD simulations and SuperSalt molecular dynamics (SuperSalt-MD) simulations. To effectively characterize the average structure of various melts, we computed partial radial distribution functions (RDFs) for all atomic pairs. As shown in Figure~\ref{fig:rdf_eos}b, the partial RDFs from the SuperSalt-MD simulations nearly overlap with those from the AIMD simulations.
Using MgCl$_2$ as an example, we performed both AIMD and SuperSalt-MD simulations under identical conditions, such as the NVT ensemble (constant number of atoms N, volume V, and temperature T) with T = 1200 K and at equilibrium volume. After a 10 ps equilibration, the average RDFs were computed from MD trajectories over 50 ps for AIMD and 100 ps for SuperSalt-MD. As shown in Figure~\ref{fig:rdf_eos}b, the partial RDFs from the SuperSalt-MD simulations (solid line) closely overlap with those from AIMD (circle dot), with both methods showing the same first and second peak positions ($R^{Na-Na}_1=3.9$ \AA; $R^{Na-Cl}_1=2.7$ \AA; $R^{Cl-Cl}_1=4.0$ \AA; $R^{Na-Na}_2=8.1$ \AA; $R^{Na-Cl}_2=6.2$ \AA; $R^{Cl-Cl}_2=8.0$ \AA) and overlapping shoulders after the first peak in both Na-Na and Cl-Cl RDFs. This indicates excellent agreement between SuperSalt-MD and AIMD structures. These results also show good agreement with other MLIP studies conducted exclusively on NaCl melts~\cite{li2021development}.
Although the atomic pair interactions are more complex in multi-component melts, both methods still exhibit the same prominent peak positions.

\subsubsection*{Density, Heat capacity, and Thermal expansion}

\begin{figure}[htbp]
    \centering
    \includegraphics[width=\textwidth]{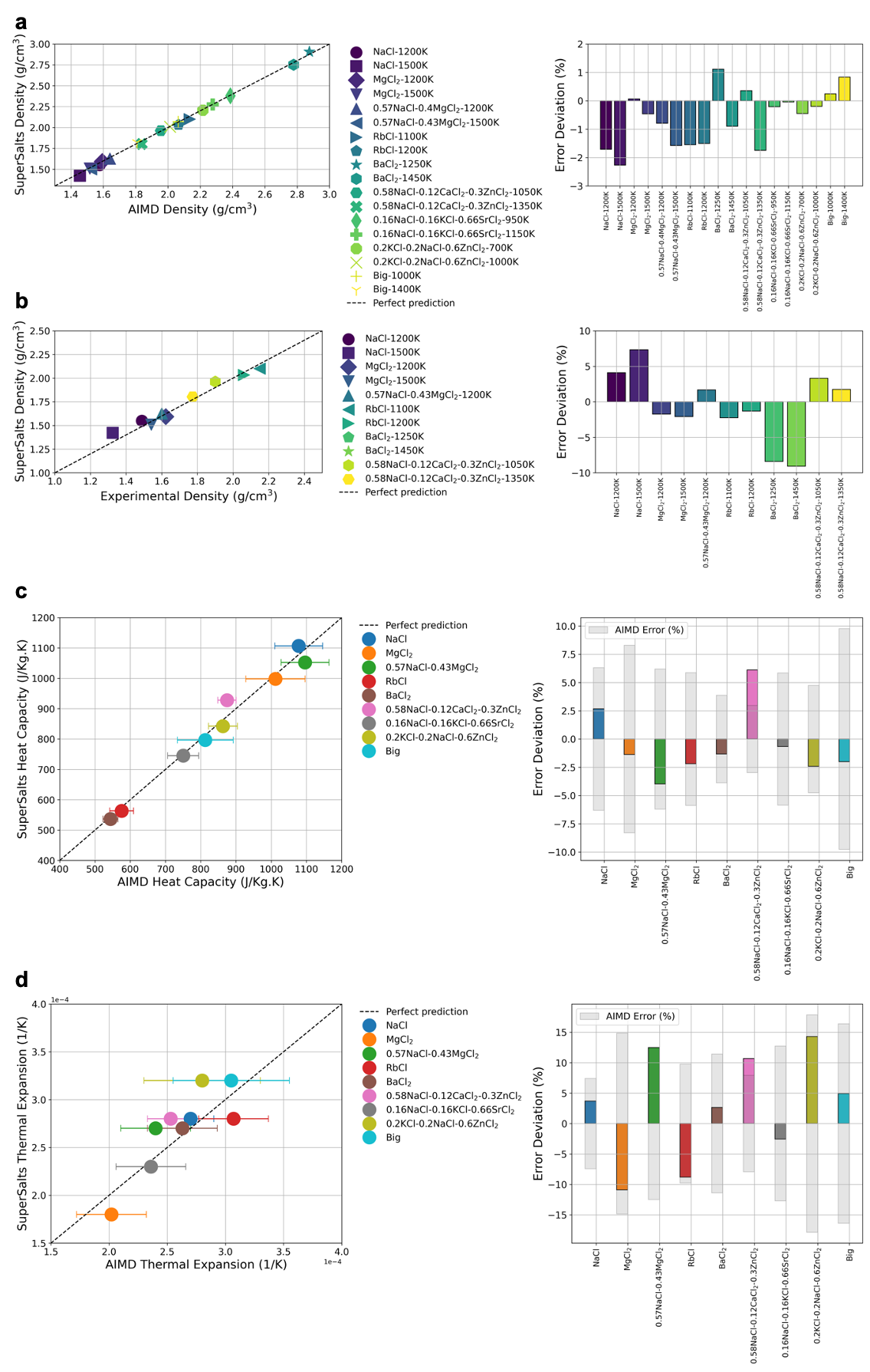}
    \caption{In the left panels, various properties calculated by SuperSalt-MD are compared with AIMD or experimental results, while in the right panels, relative error deviations are shown as histograms. (a) SuperSalt potential predictions for 9 molten salt systems at two different temperatures show excellent agreement with AIMD results, with an average deviation of less than 2\%. (b) SuperSalt potential predictions for 6 molten salt systems at different temperatures agree well with experimental results, with an average deviation of less than 5\%. (c) Heat capacities of 9 molten salts were calculated from the enthalpy deviations between two temperatures. A 100-ps-long SuperSalt-MD simulation was performed at each temperature. The largest difference between SuperSalt predictions and AIMD calculations is around 5\% for the 0.58NaCl-0.12CaCl$_2$-0.3ZnCl$_2$ system. (d) The volumetric thermal expansion coefficients for 9 molten salt systems, calculated from the volume change deviations between two temperatures, are compared with AIMD results. Error bars in AIMD calculations, corresponding to standard deviations, are shown in the left panels and gray histograms in the right panels. The 'Big' system means an 11-component system, namely, Li$_2$Na$_4$Mg$_3$K$_5$Ca$_9$ZnRb$_3$Sr$_2$Zr$_6$Cs$_2$Ba$_{2}$Cl$_{74}$
    }
    \label{density}
\end{figure}

We further compare the mass densities of different melts at various temperatures from AIMD and the SuperSalt potential, as these reflect changes in system volume and are closely linked to the local structures. To showcase the versatility and accuracy of the SuperSalt potential across a diverse range of compositions and temperatures, we selected several systems that encompass unary, binary, ternary, and multicomponent molten salts. These systems span a wide range of temperature conditions, including well-characterized salts such as NaCl, eutectic NaCl-MgCl$_2$, 0.58NaCl-0.12CaCl$_2$-0.3ZnCl$_2$, as well as more complex mixtures containing all the relevant elements (Li$_2$Na$_4$Mg$_3$K$_5$Ca$_9$ZnRb$_3$Sr$_2$Zr$_6$Cs$_2$Ba$_{2}$Cl$_{74}$).
As shown in Figure~\ref{density}a, SuperSalt demonstrates excellent agreement with AIMD, with an average deviation of less than 2\% across a wide temperature range (900 K–1400 K).
It should be noted that since both the SuperSalt potential and AIMD predictions use around 100 atoms, the density deviations are unlikely to be caused by finite-size effects. 
Compared to experimental measurements~\cite{haynes2012density,wang2017molecular,li2017survey}, as shown in Figure~\ref{density}b, the mass densities predicted by SuperSalt show a slight deviation, with an average difference of 5\%.
We believe that the density deviations observed in SuperSalt-MD simulations compared to experimental measurements may be attributed to the approximations inherent in DFT calculations used for fitting the potential, such as the choice of electronic exchange-correlation functionals and dispersion corrections. Several previous studies~\cite{nam2014first,liu2014solubility, andersson2022ab} have highlighted that dispersion interactions significantly influence the calculated density.

The excellent heat transfer performance of molten salts is critical for applications in molten salt reactors (MSR) and concentrated solar power (CSP). Among various properties, heat capacity (\(C_p\)) is the most significant factor influencing the heat transfer efficiency of liquid coolants and heat storage media. Figure~\ref{density} presents both the AIMD and SuperSalt-predicted average \(C_p\). 
These values are calculated by $C_p=(\frac{\delta H}{\delta T})_p$ based on the slope of average enthalpy $H$ at each temperature between two data points.
As demonstrated, our SuperSalt predictions for average \(C_p\) are generally in good agreement with the AIMD results, with the largest deviation being 4.8\% for 0.58NaCl-0.12CaCl$_2$-0.3ZnCl$_2$, which remains within the error margin of the AIMD data.
From the equilibrium volumes at two closely spaced temperature values, we can further estimate the volumetric thermal expansion coefficient $\alpha_V=\frac{1}{V}(\frac{\delta V}{\delta T})_p$.
Compared to AIMD results, as shown in Figure~\ref{density}c, the thermal expansion coefficients predicted by SuperSalt show a slight deviation, with an average difference of 6.7\%. This is expected since the densities have errors up to 2\%, and the error will propagate to the thermal expansion coefficient calculations, resulting in higher errors.

\subsection*{Expediting the discovery of the desired salt system}
With the highly accurate and robust SuperSalt potential, we can now perform molecular dynamics simulations with large supercells on nanosecond timescales, providing an unprecedented opportunity to explore the complex landscape of properties for molten salts.
However, calculating complex properties remains highly time-consuming for specific applications, even using MLIP in molecular dynamics simulations. For example, computing viscosity for a large supercell can take several months, greatly restricting our ability to explore uncharted regions of the melt space with desired properties.
Considering the 11-component system modeled here poses a vast compositional space (approximately forty-six trillion combinations with 1 at.\% interval), which is impractical to explore exhaustively, even for theoretical calculations.
Therefore, an efficient search algorithm is required with a proper balance between exploration and exploitation, especially when a property is difficult to calculate.

Here, the practical application of Bayesian Optimization (BO) for efficiently searching the 11-cation chloride melts space to identify compositions with optimal properties is demonstrated. Two experiments were conducted to evaluate the capabilities of the SuperSalt-BO algorithm within this 11-component molten salt system.
The first test aimed to find the composition with the highest density at 1200 K, while the second test targeted identifying a quaternary system with a density within a narrow range of 2.200–2.2001 g/cm$^3$ at 1200 K.

In the first test, where the maximum density is unknown, we used a stopping criterion based on both data and BO convergence. On the data side, the experiment halts if, for two consecutive iterations, the algorithm fails to find a better solution, and the standard deviation of the candidates in each iteration consistently decreases. On the BO side, if the surrogate model repeatedly predicts that the current solution is the best after exploring all candidates, which are generated during this iteration, and training data points, we stop the experiment.

\begin{figure}[htbp]
    \centering
    \includegraphics[width=1\textwidth]{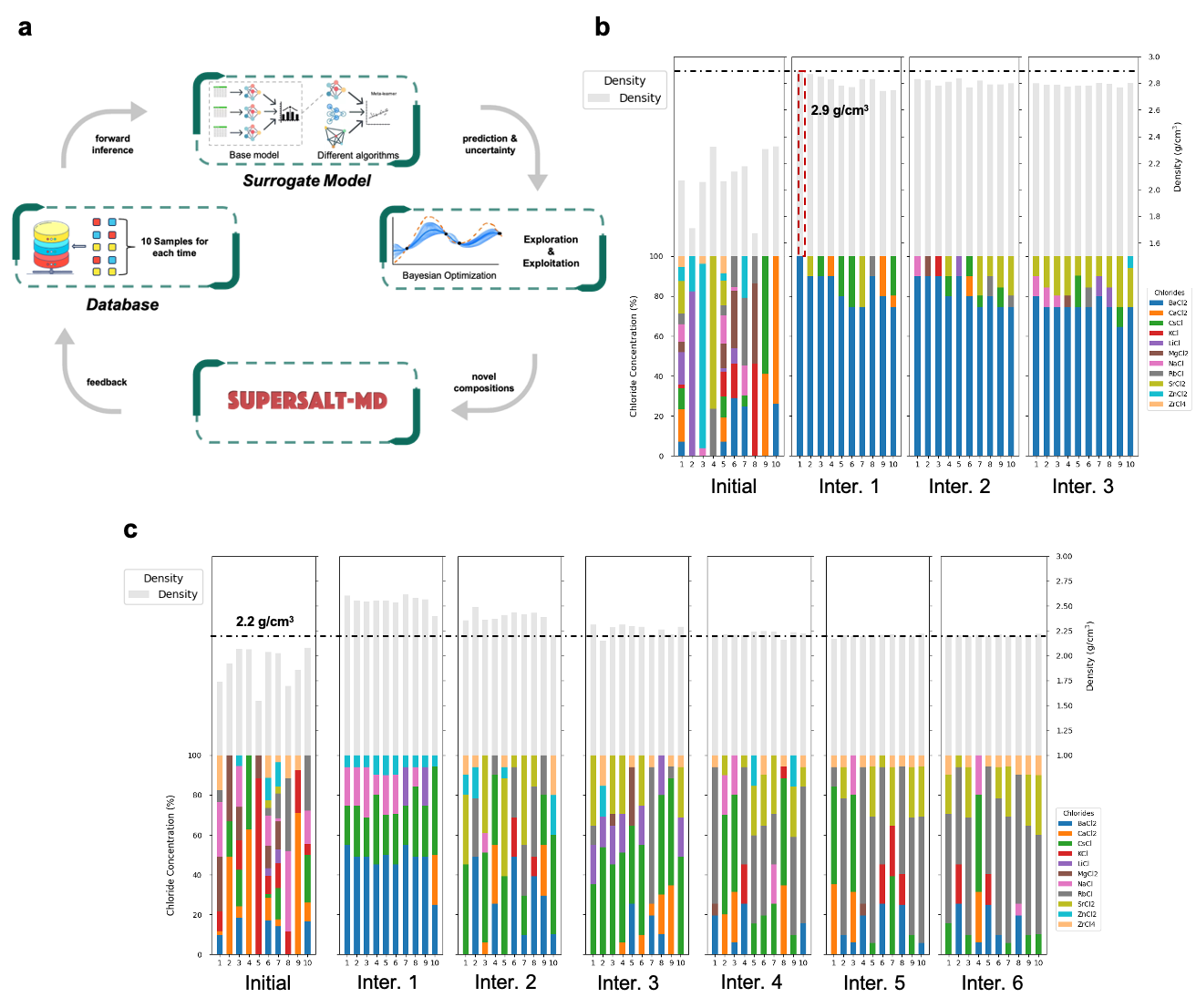}
    \caption{(a) Illustration of the workflow of SuperSalt-BO for targeting properties in 11-cation chloride melts. (b) \textbf{Target one} is to search for the composition with the largest density at 1200 K in 11-cation chloride melts. The histograms show the evolution of density with composition at 1200 K. The lower colorful bars represent the compositional distributions across iterations of the algorithm, while the upper gray bars show the corresponding densities calculated by SuperSalt-MD. (c) \textbf{Target two} is to identify a quaternary system with a density close to 2.2 g/cm$^3$ at 1200 K in this 11-cation chloride salt system. The initial datasets, consisting of 10 random compositions for both targets, are also shown for reference.
}
    \label{fig:BO}
\end{figure}

To evaluate the effectiveness of the algorithm, the Bayesian Optimization (BO) process was initiated with a deliberately poor-quality dataset, consisting of 10 randomly selected compositions with densities below 2.4 g/cm$^3$ from the previously collected data. The BO algorithm quickly found the maximum density of 2.9 g/cm$^3$ in the first iteration, corresponding to pure BaCl$_2$, as shown in Figure~\ref{fig:BO}b. Although no higher values were found in subsequent rounds, triggering the stopping criteria, we observed that the mean density of selected candidates increased steadily while their standard deviation shrank, indicating the BO's convergence. The observation that BaCl$_2$ has the highest density among the 11 components can be explained by BaCl$_2$'s relatively high atomic mass and compact ionic structure. This result confirms the effectiveness of our algorithm and strategy in efficiently navigating the molten salt property space. 

In the second test, our goal was more demanding: finding a composition within the density window of 2.200–2.2001 g/cm$^3$. Unlike the first test, which sought the global maximum, this test required the algorithm to find a precise target in a quaternary system. Starting again with 10 randomly selected points below 2.1 g/cm$^3$, our BO found a composition with a density of 2.20003 g/cm$^3$ after six iterations, corresponding to the Cs$_8$Rb$_{28}$Sr$_{10}$Zr$_5$Cl$_{76}$ composition, as shown in Figure~\ref{fig:BO}c. From the third iteration onward, our BO identified compositions with densities very close to the 2.2 g/cm$^3$ target, and by the fifth iteration, all selected candidates converged tightly around this value. This demonstrates the algorithm's capability to explore the molten salt property space with high precision. 

Interestingly, in the first iteration, the algorithm selected compositions with much higher densities (around 2.5 g/cm$^3$). This behavior can be attributed to the initial training set, where no sample had a density near 2.2 g/cm$^3$. As a result, the algorithm initially prioritized exploration by selecting compositions at the opposite end of the target density to balance the training data distribution. This ensured that the surrogate model could more robustly map the relationship between the property and the feature space. In the second iteration, the algorithm adopted a more conservative strategy, slightly lowering the mean of the overall predictions to around 2.4 g/cm$^3$. However, it preserved the exploratory tendencies of some surrogate models, which discovered a component with a density very close to the target density window. This discovery served as a turning point, allowing the algorithm to rapidly narrow down the search in subsequent iterations and converge efficiently on the target window. This adaptive process showcases the algorithm's exceptional ability to balance exploration and exploitation, dynamically adjusting its search behavior based on real-time data feedback to reach the desired outcome.

Overall, these tests illustrate that our SuperSalt-BO framework, coupled with the SuperSalt-MD calculations, provides a highly effective approach for exploring and optimizing molten salt properties in an extensive compositional space. Whether targeting a global maximum or a specific range, our model demonstrates remarkable efficiency in identifying compositions with desired properties in this 11-component molten salt system.

\section{Discussion}\label{sec13}
In this work, we have developed a comprehensive workflow for generating an efficient dataset of quantum mechanical reference data tailored to molten salt systems. Given that the behavior of salts is predominantly governed by simple pairwise electrostatics, we anticipated that fitting the 1- and 2-component systems would capture much of the essential physics. However, recognizing the potential importance of many-component interactions and the need for the MLIP to handle complex environments, we also incorporated data from multicomponent systems, including 11-cation chloride salts, to ensure transferability and accuracy across diverse compositions. Therefore, we created a database consisting of active-learned, representative 1-component, 2-component, and 11-component structures. This database enables the development of a general-purpose interatomic potential, the SuperSalt potential, applicable to the LiCl-NaCl-KCl-RbCl-CsCl-MgCl$_2$-CaCl$_2$-SrCl$_2$-BaCl$_2$-ZnCl$_2$-ZrCl$_4$ salt system across a wide range of compositions.

By minimizing a loss function that considers both potential energy and atomic forces, our SuperSalt potential achieves DFT-level accuracy in energy and force predictions while offering excellent stability for long timescale and large-length scale MD simulations under various ensembles. Extensive validation of material properties, including density, thermal expansion coefficient, specific heat, bulk modulus, and radial distribution function across various molten salt systems, shows excellent agreement between SuperSalt-MD and AIMD simulations. Furthermore, property evaluations based on SuperSalt-MD also align well with a wide range of experimental measurements, highlighting the potential of SuperSalt for robust molten salt modeling.

We demonstrate that SuperSalt-MD calculations and a SuperSalt-BO search method can effectively explore the complex landscape of molten salts and find compositions with optimal properties. The results on density show that with just tens of SuperSalt-MD calculations, a composition with the target property values can be identified. This result suggests that other properties accessible through SuperSalt-MD can be designed efficiently, accelerating the identification and optimization of suitable salt systems for various applications.

We view the present database and SuperSalt potential model as a starting point for broader studies within the molten salts community. Our efficient database generation strategy enables straightforward extensions to more complex systems. For example, to train a new model for a 12-component system, one could simply add one extra unary component, 12 relative binary systems, and active-learned 12-component configurations into our current database to ensure adequate representation—though determining the right combination for higher-order systems may require careful selection. One could also develop faster, more targeted potentials by fitting to subsets of our training data.

In the future, we plan to expand the potential to include more components, such as impurities from fission or corrosion products and F anions. Additionally, we will upgrade our SuperSalt-BO algorithm to support multi-objective, multi-task, and multi-fidelity searches. Thus, SuperSalt-MD modeling and SuperSalt-BO searching offer a promising approach to mitigating the long-standing trade-off between accuracy and computational cost in molten salt assessments. 

\section{Methods}\label{sec13}
\subsection*{Machine-learning potential fitting
}
All models trained in the work use the MACE architecture~\cite{batatia2022mace}.
MACE is an equivariant message-passing graph tensor network, where each layer encodes many-body information of atomic geometry. At each layer, many-body messages are formed using a linear combination of a tensor product basis~\cite{witt2023acepotentials,batatia2022design,darby2023tensor}. This basis is constructed by taking tensor products of a sum of two-body permutation-invariant polynomials, expanded on a spherical basis. The final output is the energy contribution of each atom to the total potential energy. For a more detailed description of the architecture, see Refs. ~\cite{batatia2022mace,magduau2023machine}.
All models referred to in this work use two MACE layers, a spherical expansion up to $l_{\text{max}} = 3$, and 4-body messages in each layer (correlation order 3). Each model uses a 64-channel dimension for tensor decomposition. We apply a radial cutoff of 8 \AA~and expand the interatomic distances into 10 Bessel functions, multiplied by a smooth polynomial cutoff function to construct radial features, which are then fed into a fully connected feed-forward neural network with three hidden layers of 64 units and SiLU non-linearities.
The irreducible representations of the messages have alternating parity, 128x0e + 128x1o for our model. 
The models are trained with the AMSGrad variant of Adam with default parameters $\beta_1$ = 0.9, $\beta_2$ = 0.999, and $\epsilon$ = 10$^{-8}$. We use a learning rate of 0.001 and an exponential moving average (EMA) learning scheduler with a decaying factor of 0.995. We employ a gradient clipping of 100.
We trained the model using a single NVIDIA A100 GPU with 80GB of RAM, and the whole process took roughly 96 GPU hours.

\subsection*{DFT computations}
The DFT calculations were performed to obtain energies, forces, and stresses of these atomic configurations using VASP 6.4.2 code~\cite{kresse1993ab}. 
The PAW-PBE potentials~\cite{perdew1996generalized,blochl1994projector,kresse1999ultrasoft} which were used in this study are Li\_sv (1s$^2$2s$^1$), Na\_pv (2p$^6$3s$^1$), Mg (3s$^2$), Cl(3s$^2$3p$^5$), K\_sv(3s$^2$3p$^6$4s$^1$), Ca\_sv(3s$^2$3p$^6$4s$^2$), Zn(4s$^2$3d$^{10}$), Rb\_sv(4s$^2$4p$^6$5s$^1$), Sr\_sv(4s$^2$4p$^6$5s$^2$), Zr\_sv(4s$^2$4p$^6$4d$^2$5s$^2$), Cs\_sv(5s$^2$5p$^6$6s$^1$), Ba\_sv(5s$^2$5p$^6$6s$^2$). To correct for dispersion effects, we used the PBE-D3 method~\cite{grimme2010consistent}. An energy cutoff of 700 eV and a K-point mesh of 1×1×1 were used to perform the DFT calculations.  

\subsection*{AIMD workflows}
To compare the performance of the potentials with DFT, we conducted AIMD simulations for several salts at different temperatures under an ambient condition. Each simulation was run for 50~ps with a time step of 1~fs using the PBE-D3 functional in VASP. The simulation cells contained around 100 atoms. From these simulations, we calculated key properties such as the radial distribution function (RDF), density, average specific heat, and average thermal expansion coefficient over the temperature range. 

\subsection*{MD workflows}
The atomic configurations we used for training were generated using MD simulations with MACE-MP0 universal potential and LAMMPS~\cite{thompson2022lammps}. Such a universal potential, which is fitted on crystalline systems, may not be accurate enough for predicting thermophysical properties of molten systems (we found significant errors when its predictions were compared to equivalent AIMD results in supplementary materials), but it is very useful for raw data generation (atomic configuration for which the energies and forces are yet to be calculated by DFT). We separately ran MD simulations for unary, binary, and 11-cation systems in the temperature range 0 K $<$ T $<$ 1600 K and in the pressure range 0 GPa $<$ P $<$ 1 GPa, each for 0.5 ns. MD simulation for each unary system started from its crystalline structure with a supercell of around 100 atoms. For each binary system, there are infinite possible compositions of A$_x$B$_{(1-x)}$, A and B being different unary salts. In previous work~\cite{attarian2025best}, we showed that only including data from unaries (A and B salts) and binaries at x = {0.33, 0.67} can provide enough data diversity to train a compositionally transferable potential across the A$_x$B$_{(1-x)}$ system. In this work, we decided to include binary data at three compositions x= {0.25, 0.50, 0.75} to add extra diversity to training data. As there are 55 possible binary combinations for an 11-cation set, overall, 55×3=165 MD simulations were run for binary systems. The initial configuration for each binary system was generated randomly using PACKMOL code~\cite{martinez2009packmol}, and each configuration had around 100 atoms. We also ran MD simulations of 40 different compositions of 11-cation salt to provide data for the multispecies system to our potential. The initial configurations for these 11-cation systems were generated using PACKMOL, and each system contained around 200 atoms. We employed the Dirichlet distribution method~\cite{ng2011dirichlet} to control and generate the compositions of the 11-component system systematically. The Dirichlet distribution, a multivariate generalization of the Beta distribution, allows for the random generation of non-negative fractions that sum to one, ensuring the proper representation of the relative proportions of each component in the system. Overall, more than 20,000,000 raw atomic configurations were generated. Using a clustering algorithm based on the HDBSCAN method, we collected around 70,000 configurations for training the potentials (roughly 1/3 unaries, 1/3 binaries, and 1/3 11-cation systems). A validation set including around 6000 configurations was also collected to be used during the training of the potential.
To test the potential, we generated two separate testing sets. The Test\_1 set included configurations from ternary salts. For an 11-cation set, there are 165 possible ternary systems. We only used A$_{0.33}B_{0.33}C_{0.34}$ compositions for each ternary system (A, B and C are different unary salts such as (NaCl)$_{0.33}$(ZnCl$_4$)$_{0.33}$(RbCl$_2$)$_{0.34}$) and at each composition we used 20 configurations (overall 3300 configurations in Test\_1). The Test\_2 set included 800 randomly generated multicomponent salts (from binary systems to 11-cation systems) with random compositions. To collect the data for each of these testing sets, the initial configurations were generated by PACKMOL, and we ran an MD simulation similar to what was done for generating training data. In the end, we randomly picked atomic configurations from the MD trajectory and later calculated the energy and force using DFT. 

\subsection*{BO workflows}
In our application of Bayesian Optimization (BO) to the problem, we explore the complete composition search space, which includes LiCl, NaCl, KCl, RbCl, CsCl, MgCl$_2$, CaCl$_2$, SrCl$_2$, BaCl$_2$, ZnCl$_2$, and ZrCl$_4$. Each component’s concentration can vary from 0\% to 100\% with an interval of 1\%, resulting in a vast search space of 46,897,636,623,981 unique compositions.

In our BO process, initially, 10 random compositions are selected as the starting training set. In each subsequent BO iteration, the model suggests 10 promising candidate compositions that are likely to meet the target criteria. After computational evaluation, if a candidate satisfies the target condition, the process halts. If none of the suggested points satisfy the condition, the new data is added to the training set, and the BO process repeats. To further enhance the model’s ability to capture intricate relationships between components, we leveraged Magpie descriptors\cite{ward_general-purpose_2016},  which not only provide comprehensive chemical descriptors for each component but also include statistical quantities that depend on the actual composition. These descriptors form a 146-dimensional feature vector for each composition, together generating a more balanced and healthy feature space in comparison to the 11-dimensional space of compositions. This feature space allows the chemical trends of scarce elements or combinations to be captured by the model as well, thus reducing bias due to unbalanced data sets, which helps the model to identify the complex behavior of materials based on a wide range of chemical properties and to learn more general and rational distributions effectively.

However, two major challenges are introduced due to the enormous search space as well as the high dimension of the feature space, how to guarantee the model extrapolation ability given the sparse data compared to vast search space, and how to adjust the sampling strategy to ensure an efficient exploration and exploitation simultaneously. 

To address the first challenge, we implemented a custom-designed ensemble model combining bootstrapping and stacking strategies\cite{ribeiro_efficient_2022}. This ensemble framework includes models like Lasso\cite{ranstam_lasso_2018}, Ridge\cite{mcdonald_ridge_2009}, ElasticNet\cite{de_mol_elastic-net_2009}, DecisionTreeRegressor\cite{noauthor_sdtr_nodate}, RandomForest\cite{rigatti_random_2017}, SVR\cite{awad_support_2015}, MLPRegressor\cite{dutt_multilayer_2022}, GradientBoostingRegressor\cite{prettenhofer_gradient_2014}, ExtraTreesRegressor\cite{sudhamathi_ensemble_2024}, and XGBoost\cite{chen_xgboost_2016} as the base learners. The key motivation behind using such a diverse set of models is twofold: firstly, the ensemble model provides strong predictive power by blending the strengths of different models. Thus, the robustness of the prediction can be guaranteed; secondly, each base model, while ensuring robust predictions, also serves as a sampler focusing on different data characteristics. Combined with our BO framework's model inference\cite{pedregosa_scikit-learn_nodate}, this guarantees diversity in the sampling process, helping to explore various data trends and thus avoiding bias toward any single model's predictions.

As for the second challenge, even with a pre-trained surrogate model, it is computationally infeasible to evaluate all potential data points across the vast search space. Furthermore, traditional sampling techniques, such as Markov Chain Monte Carlo (MCMC), particle swarm optimization, or simulated annealing, are prone to getting trapped in local optima due to the sparse initial sampling. To overcome these limitations, we implemented a hierarchical sampling strategy as follows: First, the model predicts over a 10\% interval of the search space, generating a candidate set of approximately 184,756 points. From this set, the first 5,000 compositions identified by the BO acquisition function are selected as the first level initial inputs to the Particle Swarm Optimization (PSO) algorithm, which searches for the second level fine candidate compositions for the acquisition function. This layered approach allows the algorithm to efficiently exploit the most promising regions of the search space, mitigating the risk of being stuck in local optima.

Unlike traditional BO approaches\cite{shahriari_taking_2016}, which often use a time-decaying exploration/exploitation coefficient to balance between exploring new areas and exploiting known good regions, our method employs a fixed, small value of 0.1 for the hyperparameter of the acquisition function. This ensures that the model focuses more on exploiting more reliable predictions rather than being diverted by uncertainty into less promising areas. Instead, the exploration process in our algorithm is achieved by leveraging the diverse predictions from multiple base models, each exploring different data trends. By allowing each model to exploit its area of expertise while collectively exploring more of the search space, our approach dynamically balances efficient exploration and exploitation at every iteration.


\backmatter

\bmhead{Supplementary information}

\bmhead{Acknowledgements}
We would like to express our sincere gratitude to Dr. Jicheng Guo and Dr. Ganesh Sivaraman from Argonne National Laboratory for their valuable discussions and technical assistance. 
We gratefully acknowledge support from the Department of Energy (DOE) Office of Nuclear Energy’s (NE) Nuclear Energy University Programs (NEUP) under award \# 21-24582. This work used Bridges-2 cluster~\cite{10.1145/3437359.3465593} at Pittsburgh Supercomputing Center (PSC) and Stampede3 cluster at Texas Advanced Computing Center (TACC) through allocations MAT240071 and MAT240075 from the Advanced Cyberinfrastructure Coordination Ecosystem: Services \& Support (ACCESS)~\cite{10.1145/3569951.3597559} program, which is supported by the National Science Foundation grants \#2138259, \#2138286, \#2138307, \#2137603, and \#2138296. We also used the computational resources provided by the Center for High Throughput Computing (CHTC) at the University of Wisconsin–Madison. The authors gratefully acknowledge the computing time provided
to them on the high-performance computer Lichtenberg at the NHR Centers NHR4CES at TU Darmstadt.
\newpage
\bibliography{sn-bibliography}

\end{document}